\documentclass[a4paper,10pt]{article}

\usepackage[utf8]{inputenc}
\usepackage[english]{babel}
\usepackage{fontenc}
\usepackage[pdftex]{graphicx}
\usepackage{calligra}

\usepackage{subfigure}
\usepackage[dvips]{epsfig}
\graphicspath{{./figures/}}

\usepackage{cite}
\usepackage{amsmath}
\usepackage{amsfonts}
\usepackage{amssymb}
\usepackage{textcomp}
\usepackage{xcolor}

\usepackage{authblk}

\title{Analysis of the velocity field of granular\\ hopper flow}
\author[1,2]{Magalh\~aes, F.G.R.}
\author[3,4]{Atman, A.P.F.}
\author[1]{Moreira, J.G.}
\author[2,4,5]{Herrmann, H.J.}
\affil[1]{Departamento de F\'isica, Universidade Federal de Minas Gerais, Caixa Postal 702, 30161-970 Belo Horizonte, Brazil}
\affil[2]{Computational Physics, IfB, ETH Z\"urich, H\"onggerberg, CH-8093 Z\"urich, Switzerland}
\affil[3]{Departamento de F\'isica e Matem\'atica, Centro Federal de Educa\c{c}\~ao Tecnol\'ogica de Minas Gerais, 30510-000 Belo Horizonte, Brazil}
\affil[4]{Instituto Nacional de Ci\^encia e Tecnologia Sistemas Complexos, Brazil}
\affil[5]{Departamento de F\'isica, Universidade Federal do Cear\'a, 60451-970 Fortaleza, Cear\'a, Brazil}

\begin{document}

\maketitle

\begin{abstract}
We report the analysis of radial characteristics of the flow of granular material through a conical hopper.
The discharge is simulated for various orifice sizes and hopper opening angles.
Velocity profiles are measured along two radial lines from the hopper cone vertex: along the main axis of the cone and along its wall. 
An approximate power law dependence on the distance from the orifice is observed for both profiles, although differences between them can be noted.
In order to quantify these differences, we propose a Local Mass Flow Index that is a promising tool in the direction of a more reliable classification of the flow regimes in hoppers.
\end{abstract}
 
\section{Introduction}

The study of granular flows has attracted great interest from physicists and engineers due to its 
applicability and rich phenomenology\cite{duran,campbell2006granular,herrmann1997grains},
particularly due to the applications in the industrial and mining processes where it plays a 
crucial role\cite{cleary2002modelling}.
Moreover, in the last decades, with the increasing computational performance of codes  

based on the Discrete Element Method (DEM), there has been a new surge of 
interest on the subject with the possibility of studying large systems at the scale of real experiments.
One of the oldest and most important cases of granular flow is the discharge of grains through a bottleneck.
Efforts have been made to understand the multitude of features presented by this type of flow such as clogging\cite{zuriguel2014review,janda2008,magalhaes2014jamming}, 
density fluctuations \cite{herrmann1994}
, segregation \cite{artega1990flow,magalhaes2012segregation}, 
strong fluctuations on the forces acting on silo walls\cite{ristow1995forces,goda2005three} and the
collapse of silos during discharge \cite{PhysRevLett.WallCrush}. 
Another known property of granular discharges is that two distinct flow regimes can be observed
\textemdash$~$mass flow and funnel flow \textemdash,
depending on the material properties of the grains and walls, and mostly on the hopper geometry (grain-wall friction, opening angle, orifice size, among others)\cite{jenike1961, johansonjenike62, Jenike1967, nguyen1980funnel}. 
In engineering applications, mass flow is usually desired, since it is the flow where grains leave
the hopper in a ``first-in, first-out'' order. Oppositely, funnel flow is characterized by the 
presence of stagnation zones, so that part of the grains is slower or immobile and acts as a funnel next to 
the hopper walls. In some applications, such as typically the agricultural ones, the stagnant material may deteriorate
causing direct loss of product as well as efficiency.
Therefore, the better understanding of these phenomena, particularly the influence of the design of silos, can impact directly the industry of bulk solids.

The flow regime classification is often based on visualization of the steady state. Through the decades it has evolved from the sole visual observation of stagnation zones on experiments \cite{nguyen1980funnel} to the time averaged velocity spatial distributions, either by color scaled two-dimensional fields or profile curves along a certain linear path. Velocity profiles, particularly, are widely applied in the field of granular discharge flow\cite{gonzalez2012three, medina1998velocity ,kondic2014simulations, kondic2006velocity}. The funnel flow regime is then recognized by the presence of stagnation zones in the velocity fields or a large difference between the velocity in the central region and that close to the walls. In order to quantify this difference, an index is defined\cite{jenike1961,ketterhagen2009predicting}, the Mass Flow Index ($I_ \text{MF}$), corresponding to the ratio of the average particle velocity next to the hopper wall, $ v_{\text w} $,
to the average particle velocity at the hopper centerline, $ v_{\text c} $, that is, 
$I_ \text{MF} ={ v_{\text w}  }/{ v_{\text c} }.$ It is defined such that if $I_ \text{MF} < 0.3$ the regime is
funnel flow, and mass flow otherwise. 

Ketterhagen \textit{et al.} \cite{ketterhagen2009predicting} used this index to create design charts for wedge shaped and conical
hoppers, that is, they measured $I_ \text{MF}$ from simulations for different pairs of grain-wall friction and hopper wall angle and plotted them in
a phase diagram to serve as a guideline for the designing of hoppers.
For this, they averaged $ v_{\text w}$ and $ v_{\text c}$ over a large region spanning along the hopper walls and the centerline, respectively.

The velocity dependence on the vertical coordinate is usually assumed to be equal to that of an incompressible fluid\cite{Nedderman19881507, jenike1961}.
In the case of an incompressible fluid flowing through a conical hopper with an angle $\alpha$ with the horizontal plane, we can easily obtain the velocity dependence on the vertical coordinate. 
Being the volume flow rate defined as $Q=A  v$, where $A$ is the cross sectional area and $v$ is the velocity. The flow rate must be constant along the direction of the flow, provided that the density is homogeneous.
Thus, consider a point at height $z_o$ on the hopper opening, where $Q_o=A_o v_o$, and an arbitrary point at height $z$
in the hopper, where $Q=A  v$, then
$$ \frac{Q}{Q_{o}}=\frac{ Av}{A_{o} v_{o}} =\frac{ \pi \left(z \tan \alpha \right)^2 v}{\pi \left(z_o \tan \alpha \right)^2 v_o}=1$$ and, from that, 
\begin{equation}
v = \left( \frac{z}{z_o} \right) ^{-2} v_o .
\label{eq:vpropz2}
\end{equation}
Therefore, for an incompressible fluid, the velocity decays as a power law along $z$ with exponent $2$. 
However, if Equation \ref{eq:vpropz2} holds for a granular discharge, the procedure used by Ketterhagen \textit{et al.} \cite{ketterhagen2009predicting} may lead to an $I_ \text{MF}$ value that is highly sensitive on position and size of the averaging region. This would cause their design charts to be applicable only if the same specifications were used. In fact, Gonz\'alez-Montellano \textit{et al.} \cite{gonzalez2012three} calculated the ratio of the wall and center velocities, but averaging them over small regions at a few different heights along a square cross-section silo with a hopper-bottom that also has square cross-section. Although the ratio is approximately equal to one all along the vertical silo, they observed that inside the hopper the ratio largely fluctuates reaching values below $0.3$ at some points, even though they consider the discharge to be in the mass flow regime.

Granular flow may differ from incompressible fluids as there are voids and, therefore, density heterogeneities occur, even though each grain is roughly incompressible by itself. In fact, one can expect that a larger density would be observed at positions farther from the hopper outlet, where the velocity is smaller and the grains are in a densely packed structure, as compared to regions closer to the orifice, where the velocity is larger and grains are looser. As shown by Sielamowicz and Czech\cite{sielamowicz2010analysis} in experiments discharging amaranth seeds through a two dimensional hopper, the assumption of an incompressible fluid is not always valid. They found discrepancies from the 2D equivalent to Equation \ref{eq:vpropz2}, specially, close to the outlet. Additionally, their flow regime was closer to funnel flow as there is the formation of a channel of flowing grains in the central region of the hopper, while the grains close to the walls of the hopper are stagnant.

In order to contribute to this discussion, we present in this paper the velocity profiles along the centerline and along the walls of a conical hopper and we show that they behave roughly as power laws. Nevertheless, there are some differences between $v_{\text c}$ and $v_{\text w}$ profiles that carry information about the flow regime. We propose the analysis of the Local Mass Flow Index profile as a method capable of extracting that information in a clear way. Also, this method should be applicable independently of the validity of any assumption on the velocity radial dependence. Furthermore, we believe that the Local Mass Flow Index analysis may lead to a more precise definition of the flow regime boundaries.

\section{Model}
The microscopic model used for the grain interactions was described in details by Pöschel and Schwager
\cite{PoschelBook}. A particle $i$, with radius $R_i$ and at position $\vec r_i$, is in contact with a
particle $j$ if $\xi = \left[ (R_i+R_j) -  \left| \vec r_{j} - \vec{r_{i}} \right| \right] > 0$, where
$\xi$ is the overlap length.
The normal force $f_n$ is given by the Hertzian Law:
\begin{equation}
 f_n = E_\text{eff} \sqrt{R_\text{eff}} \sqrt{ \xi } \left( \xi + A \dot \xi  \right) , 
\end{equation}
where $\dot \xi$ is the normal relative velocity; $E_\text{eff}=(E_i^{-1}+ E_j^{-1})^{-1} $ is the effective
elastic coefficient; $E_i = (4/3)[Y_i/(1-\nu_i^2)]$, where $Y_i$ is the Young modulus and $\nu_i$ is the Poisson
ratio of the material of particle $i$; $R_\text{eff}=(R_i^{-1}+ R_j^{-1})^{-1} $ is the effective radius and $A$ is
a normal damping constant \cite{brilliantov1996model}. We also impose positive values for $f_n$, so that unphysical
attractive forces are avoided.
Therefore, the normal force vector is
\begin{equation}
\vec F_n = \text{max} \left( 0, f_n \right)  {\hat e}_n ,
\label{eq:nforce}
\end{equation}
where ${\hat e}_n = \left( \vec r_{j} - \vec{r_{i}} \right)/\left| \vec r_{j} - \vec{r_{i}} \right|$.

The tangential force is implemented in the spirit of Cundall and Strack's seminal paper \cite{CundallStrack}.
In the static regime the force is represented by an elastic term, analogously to a spring being attached to the
points of the first contact of each of the particles surface, and a dissipative term, 

\begin{equation} 
\vec f_0 = -k_t \vec \zeta  - \gamma_t \vec v_t
\end{equation}
where $\vec \zeta$ is the spring displacement, $k_t$ is the spring stiffness, 
$\gamma_t$ is a tangential damping constant and 
$\vec v_t = \vec v_{ij}- {\hat e}_n ({\hat e}_n \cdot \vec v_{ij} ) $ is the tangential velocity. Only the tangential displacements are considered, so that $\vec \zeta$ and $\vec v_t$ are parallel to each other.
If the tangential force is larger than the static friction limit, that is $f_0 > \mu_s f_n$, then the particles
are in the dynamic regime, where the force is proportional to the normal force,
\begin{equation}
f_c = \mu_d f_n,
\end{equation}
being $\mu_s$ and $\mu_d$, respectively, the coefficient of static and of dynamic friction.
Therefore, the tangential force is

\begin{equation}
   \vec F_t  = \hat e _t \cdot \left\{  
   \begin{array}[]{l l}
   f_0 & \quad \text{if $f_0 \leq \mu_s f_n$ (static regime)}\\
   f_c & \quad \text{if $f_0 > \mu_s f_n$ (dynamic regime)}
   \end{array}                                                                                                                                                                                           \right. 
\label{eq:tforce}
\end{equation}

We use material parameters as follows:
$E= 10^8  \text{ Pa}$,
corresponding to $Y \sim 10^8 \text{ Pa}$ and $\nu \sim 10^{-1}$, $\mu_s=0.8$, $\mu_d=0.6$, $k_t=100 \text{ N}/\text{m}$,
$\gamma_t=0.1 \text{ Kg}/\text{s}$, $A=0.01 \text{ s}$. In order to avoid crystallization, the radii of the
spheres are randomly chosen from a uniform distribution from an interval of $5\%$ deviation around $R = 5  \text{ mm}$. The density is
$8\times 10^{3} \text{ kg}/\text{m}^3$, therefore each grain weights about $4.2 \text{g}$. Additionally to the contact
forces, gravity acts on each particle and the acceleration of gravity in the simulations is set to
$g=10 \text{ m/s}^2$ in the direction of $-\hat{z}$. The integration of Newton's equations of motion was
performed using a fourth order Gear's predictor-corrector algorithm\cite{PoschelBook}. The time step was chosen as
$\Delta t=  10^{-6}s$ based on previous works\cite{ristow1995forces}.

In our simulations we use conical hoppers with 
different tilt angles 
$\alpha= 45^{\circ} \text{, } 60 ^{\circ} \text{ or } 75 ^{\circ}$,
measured from the horizontal plane, 
and study discharges of $20000$ grains for orifice diameters  $D= 6 R, 8 R, 12 R,16R$ and $24R $. 
The walls are simulated by the force law given by equations \ref{eq:nforce} and \ref{eq:tforce} where the overlap length $\xi$ is that of the grain with the wall surface. Additionally, $R_\text{eff}$ is chosen to be equal to the radius of the grain. The other material parameters are equal to those of collisions between grains.

The system is initialized by releasing the grains from a regular cubic lattice configuration inside the hopper with the grain centers
separated typically by $3R$ or $4R$ with small random velocities while the hopper orifice is kept closed. 
The orifice is opened after a time of typically $0.5\text{ s}$ ($5 \times 10^5$ time steps)
and then the discharge begins. 

In order to maintain a stationary flow, grains that are more than one diameter below the outlet are reinserted at randomly chosen positions at the top surface of the material in the hopper, this procedure is carried out at regular intervals of $0.1 \text{ s}$ ($10^5$ time steps).
Thus, the number of particles inside the hopper is kept approximately constant
all along the discharging process.

\begin{figure}[th]
 \center
 \includegraphics[width=8cm]{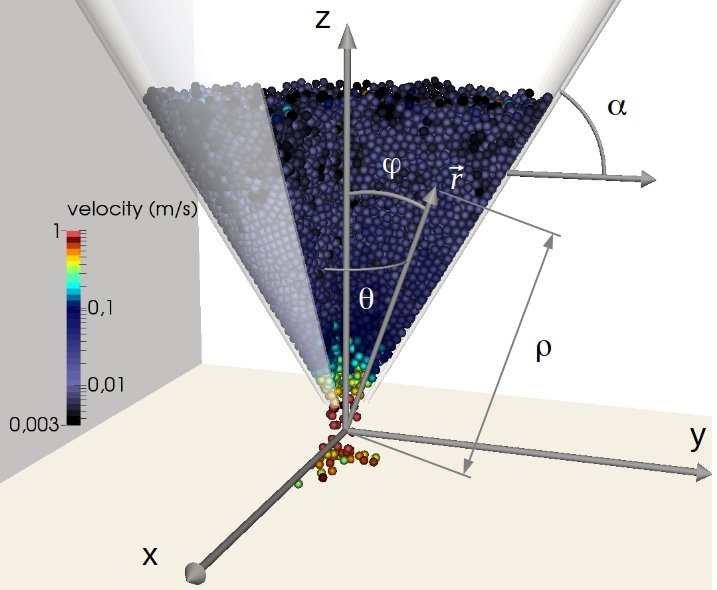}
 \caption{Snapshot of hopper and flowing grains. Cartesian and spherical coordinates are shown, respectively, as $(x,y,z)$ 
 and $(\rho, \theta, \varphi)$.}
 \label{fig:snapshot}
\end{figure}

A snapshot of the system configuration can be seen in Figure \ref{fig:snapshot}. The Cartesian and the spherical coordinate system are shown, for both cases the origin is chosen to coincide with the cone vertex. 
Also, the hopper angle with the horizontal plane, $\alpha$, is defined.

\section{Results and Discussion}
\label{sec:Results}

\begin{figure}[th]
\center
\includegraphics[width=12cm]{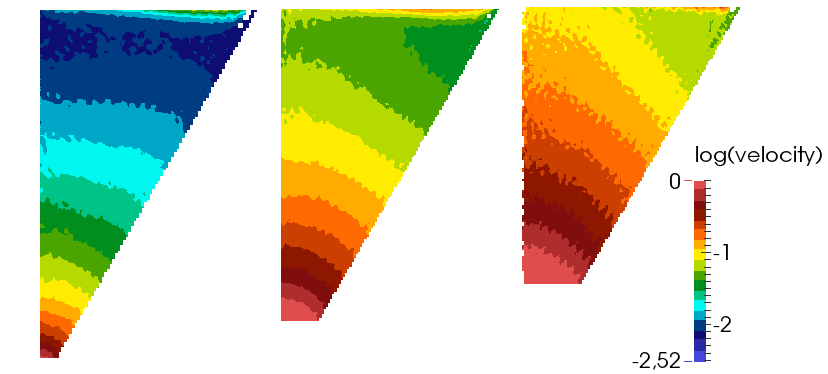}
\caption{Velocity field for $\alpha= 60^{\circ}$ and $D=8R,16R$ and $24R$, from left to right.}
\label{fig:landscape}

\end{figure}

In order to visualize the steady-state flow, we create a velocity field.
It is obtained by constructing a regular square grid over the plane $(\rho,\varphi)$ representing by a color in each cell
the modulus of the average velocity taken over grains inside that cell and the neighboring cells. That is, we average also over $\theta$ and thus, by this procedure, we exploit the azimuthal symmetry of the system. Additionally, the velocity is averaged over snapshots from the whole simulation, snapshots are taken every $0.005s$ and the duration of each simulation is typically between $5s$ and $10s$. Some of those
fields are shown in Figure \ref{fig:landscape} for a few choices of orifice widths. Note that, by increasing the
orifice size, the overall velocity increases. 
The iso-velocity curves, i.e. the borderline between two different color shades, are always crossing the entire horizontal extension of the hopper for
orifice size $D=8R$, while for $D=16R$ there are enclosed regions close to the wall. Also, the general shape of these
curves changes with the orifice size, being more horizontal for smaller orifice sizes and increasingly inclined for larger ones. This means that for the same distance from the vertex of the hopper the average velocity of the grains is larger at the center of the hopper than close to its walls.

The velocity field analysis may be useful to differentiate extreme states such as a fully developed funnel flow and a very typical mass flow. However, it is unable to quantify more subtle flow differences.
Thus, in order to better characterize the flow, we contrast the two most different regions of the hopper, that is, we compute the velocity profile along $\rho$, the radial distance from the hopper cone vertex (Figure \ref{fig:snapshot}), close to the wall of the hopper, $v_{\text w}(\rho)$, and along its central line, $v_{\text c}(\rho)$. 
Furthermore, we generalize the concept of Mass Flow Index to a Local Mass Flow Index profile, $I_ \text{LMF}(\rho)$, similarly as in  Gonz\'alez-Montellano \textit{et al.}\cite{gonzalez2012three}. It can be obtained directly from the velocity profiles as $$I_ \text{LMF} (\rho)= \frac{v_{\text w}(\rho)}{v_{\text c}(\rho)}.$$ 

\begin{figure}[h!]
 \center
      \includegraphics[height=7cm]{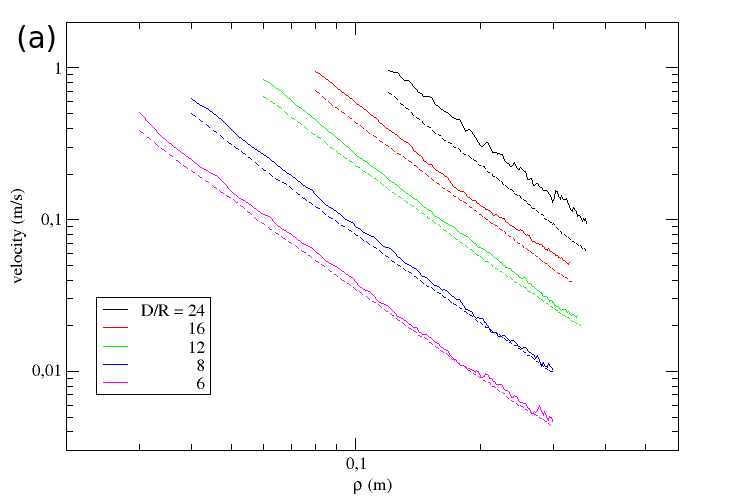} 
      \includegraphics[height=7cm]{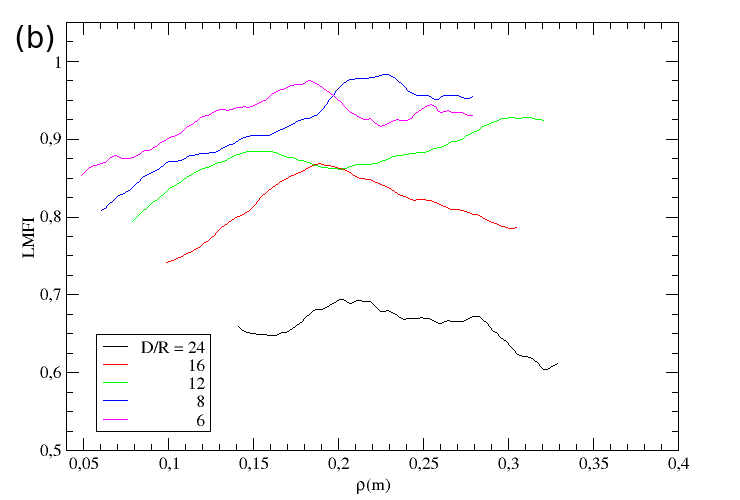}
 \caption{Velocity profiles for center (solid line) and wall (dashed line) velocities in log-log scale (a) and Local Mass Flow Index (b) for different orifice diameters and 
 $\alpha=60^\circ$. The profiles are power laws, but slightly different inclinations between $v_{\text w}$ and $v_{\text c}$ can be observed for some cases what causes the $I_{\text LMF}$ to rise. The larger the orifice diameter the larger is the gap between center and wall velocities and consequently the lower is $I_{\text LMF}$.}
 \label{fig:graph_w}
\end{figure}

We apply this analysis to data from simulations of discharges in hoppers of different orifice sizes,
Figure \ref{fig:graph_w}. 
In panel (a), a power law decay is observed for the velocity profiles along $\rho$. As discussed before, this behavior is not surprising, however, discrepancies between  $v_{\text w}$ and  $v_{\text c}$ are visible. We believe that these discrepancies carry information about the flow regime and should not be disregarded.
For small orifice sizes, namely $D=6R$ and $D= 8R$, the slope of the velocity along the center of the hopper appears to be larger than
that along the walls, until they collapse into approximately the same curve further up in the hopper. On the other hand, for large orifice
sizes, $D=16R$ and $D=24R$, there is a large gap between $v_{\text c}$ and $v_{\text w}$ all along $\rho$, and the 
slopes are very similar. Therefore, the gap is approximately maintained even for large $\rho$. One could have analyzed this also through the $I_ \text{LMF}$ in Figure \ref{fig:graph_w}b. The simulations with small opening diameters show $I_ \text{LMF}$ curves increasing with $\rho$ until they reach values close to unity, meaning the wall and center velocities are approaching identical values at higher positions. On the other hand, the curves for large opening widths fluctuate around much smaller $I_ \text{LMF}$ values, being a consequence of the approximately constant gap between $v_{\text c}$ and $v_{\text w}$. This seems to be a precursor of funnel flow where the wall velocity is much smaller than the center velocity all along the hopper.

\begin{figure}[h!]
  \centering
	\includegraphics[height=7cm]{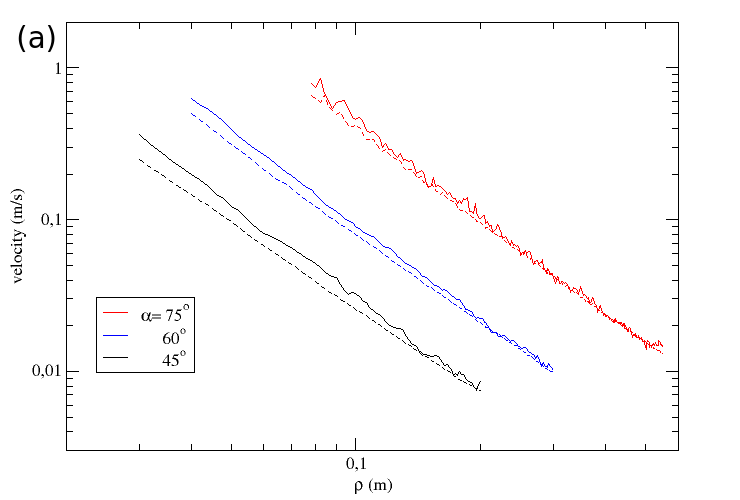} 
	\includegraphics[height=7cm]{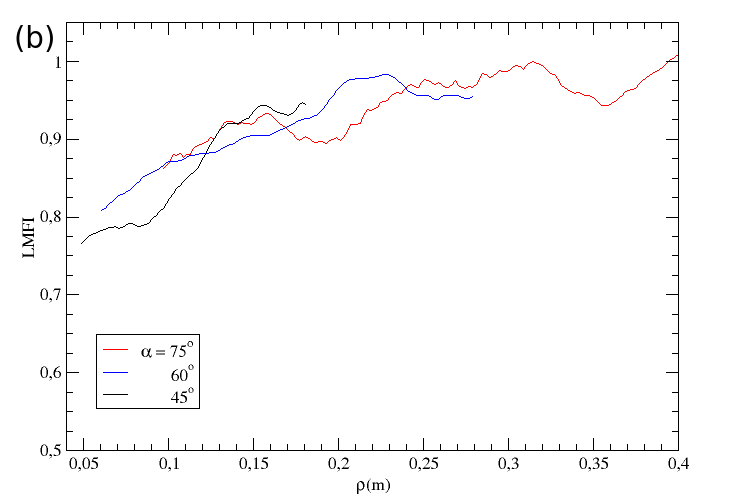}
 \caption{Velocity profiles for center (solid line) and wall (dashed line) velocities in log-log scale (a) and Local Mass Flow Index (b) for different hopper angles and $D=8R$.
 As in figure \ref{fig:graph_w}a, power law behavior is observed for the velocity profiles. Here, the differences between the center and
 wall velocities close to the outlet are larger for smaller $\theta$, however, the difference vanishes for large $\rho$ in all cases. This is represented in panel (b) by the low values of $I_{\text LMF}$ for small $\rho$ that then rise up to values close to unity.}
 \label{fig:graph_alpha}
\end{figure}

Additionally, we present the analogous velocity profiles for hoppers of different opening angles,
Figure \ref{fig:graph_alpha}a. The same general behavior is observed, that is, a power law for both velocities. Differences between $v_{\text c}$ and $v_{\text w}$ are larger close to the outlet and decrease with increasing $\rho$. 
It is interesting to note that the $I_ \text{LMF}$ curves for each angle $\alpha$, in Figure \ref{fig:graph_alpha}b, are identical within the fluctuation range. This is a very interesting effect that we intend to investigate further for other opening angles.

According to the literature \cite{Nedderman19881507, jenike1961}, the velocity radial profiles should decay as a power law with exponent equals to 2, as calculated for an incompressible fluid in the Introduction of this paper. 
For all the cases, we obtained exponents around that value, the smallest value being $1.868(3)$ for $R=12$ and $\alpha = 45^\circ$ and the largest value being $2.176(5)$ for $R=24$ and $\alpha=60^\circ$, both for the wall velocity profile.
However, we observe in Figures \ref{fig:graph_w}a and \ref{fig:graph_alpha}a that for the center velocity the slope of the profile varies along $\rho$. 
This is easily observed, for instance, for $R=12$ and $\alpha = 45^\circ$ (lower curves in Figure \ref{fig:graph_alpha}a) where the center and wall velocity profiles are very close and approximately parallel for large $\rho$, but they separate close to the orifice.
We believe this is an effect intrinsically related to the flow regime of the discharge and the local density heterogeneities.
Moreover, the differences between the central and the wall velocity profiles can be highlighted by the Local Mass Flow Index. 
In most cases, the $I_ \text{LMF}$ shows an increase with $\rho$, presenting intermediate values, around $0.8$, close to the outlet and rising further up until a value close to unity which means that $v_{\text c}$ and $v_{\text w}$ are equal. This effect is not observable by any other analysis. Also, for the case where the measured  $I_ \text{LMF}$ is approximately constant, namely $D=24R$ and $\alpha=60^\circ$, its value is relatively small, that is, close to the limit for funnel flow regime, even though it is expected to present mass flow. 

In summary, our results show that the Local Mass Flow Index profile varies along $\rho$ for many hopper angles and opening widths raising concerns over the characterization based solely on the Mass Flow Index, where this information is lost. The flow regime characterization is far too complex to be reduced to a single index. Nevertheless, the analysis of the $I_ \text{LMF}$ profiles was shown to be a very powerful tool to describe the more subtle aspects of the flow clearly. Although our simulations were limited to mass flow regime, we expect that more distinctive features will be observed for discharges in the funnel flow regime, what is part of our current investigations, which demands considerable larger CPU times.

\section{Conclusions}
\label{sec:Conclusion}

We have investigated the influence of the orifice size and hopper opening angle on the velocity fields and
profiles of spherical grains flowing through a conical hopper. 

The fields show the presence of low velocity zones close to the walls of the hopper, only observed in systems with large outlet sizes. They seem to be precursors of stagnation zones that appear in fully developed funnel flow. Also,for larger orifice widths as compared to the smaller ones studied, the iso-velocity curves present a larger inclination, that is, the average velocity at the center of the hopper is larger than the velocity close to its wall for the same distance from the hopper vertex. Although these differences do not characterize different flow regimes, they show that there is a spectrum of behaviors for each flow regime.

The velocity profiles  along the axis of the hopper and close to the walls were studied, as well as their ratio, the $I_ \text{LMF}$ profile. Both velocity profiles show approximately a power law dependence with the distance from the vertex of the hopper for the considered cases. In hoppers with large orifice widths, there is an upshift
of the velocity profile along the center of the hopper as compared to that close to the walls, but the slopes are roughly the same.
This is accompanied by a $I_ \text{LMF}$ that fluctuates around a certain value that can be as low as $0.65$ for the largest orifice width analyzed. 
On the other hand, for small outlets, the difference between the profiles decreases with the distance from the opening. This decrease is clearer when the angle of the hopper is changed. The smaller the angle of the walls of the hopper with the horizontal plane, the larger the slope of the central velocity profile compared to the
wall velocity profile. In all cases, the velocity profiles do not cross each other, so that, for a distance farther
from the orifice, the difference of the slopes becomes negligible. This decrease in difference between the velocity profiles is seen as an increase of $I_ \text{LMF}$. The $I_ \text{LMF}$ stabilizes with values close to one in the regions where the velocity profiles collapse.

We believe that the presented comparative analysis of the velocity profile at the center and close to the walls of the hopper accompanied by the $I_ \text{LMF}$ profile
is a promising tool for the classification the granular flow regimes. It provides a broader and more intelligible picture of the flow state than the techniques currently applied in the field and it lacks the set-up sensitivity of other methods. We are currently working on applying this analysis to different silo geometries, such as flat and conical bottom cylindrical silos. In those cases a fully developed funnel flow is more easily accessible and, for that regime, we expect the $I_ \text{LMF}$ analysis to present some new insights.

\textbf{Acknowledgements} This project was partially funded by CAPES, CNPq, Ci\^encias sem Fronteiras and Grant No. FP7-319968-FlowCCS of the European Research Council (ERC) Advanced Grant. FGRM would like to thank Alessandro Leonardi for helpful discussions, as well as the whole Computational Physics Group at ETH Z\"urich for the hospitality and resources.

\bibliography{artigo}
\bibliographystyle{unsrt}

\end{document}